\documentclass[showpacs,pre,aps,twocolumn,superscriptaddress]{revtex4-1}

\pdfoutput=1

\usepackage{amsmath,amsfonts,amssymb,bm,graphicx,hyperref}
\usepackage{color}

\newcommand{\eps}{\varepsilon}

\newcommand{\ee}{\mathrm{e}}

\newcommand{\CC}{\mathcal{C}}

\begin{document}

\author{Robert L. Jack}
\address{Department of Physics, University of Bath, Bath, BA2 7AY, United Kingdom}
\author{Juan P. Garrahan}
\address{School of Physics and Astronomy, University of Nottingham, Nottingham, NG7 2RD, United Kingdom}
\title{Phase transition for quenched coupled replicas in a plaquette spin model of glasses}

\begin{abstract}
We study a three-dimensional plaquette spin model whose low temperature dynamics is glassy, due to localised defects and effective kinetic constraints.  While the thermodynamics of this system is smooth at all temperatures, we show that coupling it to a second system with a fixed (quenched) configuration can lead to a phase transition, at finite coupling.  The order parameter is the overlap between the copies, and the transition is between phases of low and high overlap. We find  critical points whose properties are consistent with random-field Ising universality.  We analyse the interfacial free energy cost between the high- and low-overlap states that coexist at (and below) the critical point, and we use this cost as the basis for a finite-size scaling analysis.  We discuss these results in the context of mean-field  and dynamical facilitation theories of the glass transition. 
\end{abstract}

\maketitle

%
%

\emph{Introduction} --
There are several theories that aim to explain the rapid increase in the viscosity of supercooled liquids close to their glass transitions \cite{Ediger1996,Berthier2011,Biroli2013}.
Some of these theories predict that phase transitions should occur, either on cooling or in response to some kind
of external field~\cite{Lubchenko2007,Chandler2010,Franz1997,Hedges2009,Cammarota2012}.  
Such phase transitions are relevant because their associated order parameter fluctuations may
explain the characteristic fluctuations that are observed in the dynamics of supercooled liquids~\cite{Ediger2000}.  
The theory of dynamical faciliation (DF) \cite{Garrahan2002b,Chandler2010} 
is based on a dynamical order parameter, 
and the associated dynamical phase transitions occur in ensembles of trajectories~\cite{Merolle2005,*Jack2006,*Garrahan2007,Hedges2009,Pitard2011,*Speck2012b,*Speck2012}.  
This theory is encapsulated by a class of simple model systems that includes both kinetically constrained models~\cite{Ritort2003} and plaquette spin models~\cite{Newman1999,*Garrahan2000,*Garrahan2002}.  A different theoretical approach is based on mean-field calculations within a replica formalism~\cite{Franz1997,Franz2011}.  It proposes a static order parameter -- the {\em overlap} -- which measures the similarity between two randomly chosen configurations of the system in equilibrium.  This approach predicts a phase transition, as found in certain spin-glass models, 
that occurs when configurations of the system are biased to be similar to fixed {\em reference configurations}.  

In this paper, we consider a three-dimensional plaquette spin model -- the {square pyramid model} (SPyM)~\cite{Turner2015} -- whose relaxation behavior is well-described by
DF theory.  Using computer simulations, we show that this model also exhibits phase transitions when biased by its overlap to (quenched) reference configurations.   
The properties of the thermodynamic phase transitions that we find are consistent with the universality class of the random-field
Ising model (RFIM), as predicted by field theories for generic glassy systems~\cite{Franz2013,Biroli2014}.  Evidence for this scenario
has also been found in atomistic simulations~\cite{Cammarota2010,Berthier2015}.  
Thus, at low temperatures,
the SPyM exhibits both anomalous dynamic fluctuations (as predicted by DF) and anomalous overlap fluctuations (as predicted by
mean-field).  These results are important for two reasons: first, they provide strong evidence that theoretical predictions 
of RFIM criticality~\cite{Franz2013,Biroli2014} are applicable in three-dimensional systems; second, the fact that a single
model displays features of both mean-field theories and DF theory provides an opportunity
for a connection between these apparently contradictory scenarios~\cite{Biroli2013}.

\emph{Model} --
The SPyM~\cite{Turner2015}
consists of Ising spins $s_i=\pm 1$ on the vertices of a body-centred cubic lattice, with periodic boundaries.
We use $\CC=(s_1,\dots,s_N)$ to indicate a configuration of the system, whose energy is
$E_0=-\frac{J}{2} \sum_\mu s_{i_\mu} s_{j_\mu} s_{k_\mu} s_{l_\mu} s_{m_\mu}$ where the sum runs over upward-pointing square pyramids within the
lattice, and the five spins $s_{i_\mu} \dots s_{m_\mu}$ are on the five vertices of pyramid $\mu$.  The parameter $J$ sets the energy scale, and
the linear system size is $L$, with a total of $N=L^3$ spins.
Pyramids $\mu$ for which the product of spins is equal to $-1$ are \emph{defects} which carry an energy $J$: 
from a thermodynamic viewpoint they are non-interacting and their statistics are those of an ideal gas.  Hence for a single system
there are no thermodynamic phase transitions at any finite temperature.  However, the dynamical behavior of the defects is complex
and co-operative: by analogy with a similar model in two dimensions~\cite{Newman1999}, it is believed that for low temperatures $T$, 
the relaxation time $\tau$ diverges $\ln \tau \sim T^{-2}$ \cite{Turner2015}, and the range of
certain multi-point spin correlations also diverges~\cite{Jack2005caging}.

The overlap is $Q(\CC,\CC')=\frac1N \sum_i s_i s_i'$, which measures the similarity between spin configurations.  
We draw a configuration $\CC'$ at random from an equilibrium distribution at temperature $T'$.  Holding $\CC'$ fixed, we calculate expectation
values with respect to the distribution $p(\CC|\CC') = Z(\CC')^{-1} ~ \ee^{[\eps N Q(\CC,\CC')-E_0(\CC)]/T}$, where $Z(\CC')$ is a partition function.  
Thus for $\eps=0$ we consider equilibrium
behaviour of the isolated SPyM at temperature $T$, while increasing the coupling $\eps$ biases the configuration $\CC$ to increase its overlap with the reference $\CC'$.  Finally we perform an average over the reference
configuration $\CC'$.  We set $J=1$ which fixes our energy unit, so the dimensionless parameters of the system are $(T,T',\eps)$.
We use a Monte Carlo sampling scheme to study these coupled systems,
 as described in Appendix A. 

\emph{Results} --
Fig.~\ref{fig:bbp} shows results for $T'=T$, in which case the reference configuration is representative of thermal equilibrium.
The mean overlap
$\langle Q \rangle$ in Fig.~\ref{fig:bbp}(a) increases with $\eps$, as expected.   
This increase is gradual for small $\eps$, before a steep increase sets in at larger $\eps$.  
The theoretical prediction~\cite{Franz1997,Franz2013,Biroli2014} is that the gradient $d\langle Q\rangle/d\eps$ should diverge at an RFIM critical point at some
$(T_c,\eps_c)$, and that for $T<T_c$, one should observe (in the thermodynamic limit) a first-order transition, with a
discontinuous jump in $\langle Q \rangle$.  In these 
finite systems, no divergences are observed: to demonstrate the existence of the phase transition we use finite-size scaling methods.

The reference configuration $\CC'$ is a source of quenched disorder in this problem, and averages are calculated in two stages:
first a thermal average at fixed $\CC'$, which we denote by $\langle\cdot\rangle_{\CC'}$ and then a disorder average, denoted
by $\overline{(\cdot)}$.  The notation $\langle \cdot \rangle \equiv \overline{\langle\cdot\rangle_{\CC'}}$ indicates the double average.
Fig~\ref{fig:bbp}(b) shows the average $\langle Q \rangle$ as well as the behaviour of $\langle Q\rangle_{\CC'}$ for eight representative configurations
$\CC'$.  For each $\CC'$, one sees a very sharp jump in the overlap at a sample-dependent field $\eps^*_{\CC'}$. 
However, the jump in the average overlap is broadened
out due to fluctuations in $\eps^*_{\CC'}$ between samples.  This is the expected behaviour at an RFIM critical point~\cite{Fisher1986} 
-- it means that the disorder
is the dominant source of fluctuations in the problem.

Fig.~\ref{fig:bbp}(c) shows the (total) susceptibility $\chi_{\rm tot} = N \langle \delta Q^2\rangle = N [\langle Q^2 \rangle - \langle Q \rangle]$.
As the temperature is reduced, the fluctuations grow rapidly and depend increasingly strongly on the
system size, as expected in the vicinity of a phase transition.  Finally, Fig.~\ref{fig:bbp}(d) shows the distribution
of the overlap $Q$, evaluated at the field $\eps^*$ which maximises $\chi_{\rm tot}$.  
For the lower temperatures, a two-peaked structure is clearly visible.  Considering these data together with Fig.~\ref{fig:bbp}(b), one
 sees that typical reference configurations contribute to either the low-$Q$ or high-$Q$ peak, as expected for RFIM
criticality: the proportion of configurations $\CC'$ that contribute simultaneously to both peaks scales as $L^{-\theta}$~\cite{Fisher1986} where
$\theta$ is a critical exponent for the RFIM (in three dimensions, $\theta\approx 1.5$~\cite{vink2010,fytas2013}).  Also, $P(Q)$ in Fig.~\ref{fig:bbp}(d) has
an asymmetric shape, with a very broad high-$Q$ peak.  From Fig.~\ref{fig:bbp}(b), we attribute this broadness to the fluctuations
between reference configurations: the position and the size of the jump in $\langle Q \rangle$ vary signficantly.  Finally, 
we note that numerical uncertainties in $P(Q)$ are considerable: where the probability is small our estimate of $P(Q)$ 
may be dominated by just one or two out of the 64 reference configurations considered.  We return to this point below.

Taken together, the data in Fig.~\ref{fig:bbp} are consistent with an RFIM critical point for $T'=T$ in the range $0.3-0.4$.  
However, while the MC methods used here lead to efficient simulations even close to phase transitions (see Appendix A), the slow (glassy) relaxation of the model 
limits the temperatures and system sizes that we can consider.  If an RFIM critical point exists for $T'=T$, we expect similar critical points
at other reference temperatures $T'$. We therefore analyse a lower reference temperature $T'=0.25$, providing
further evidence for RFIM criticality in this system.  

\begin{figure}
\includegraphics[width=8.5cm]{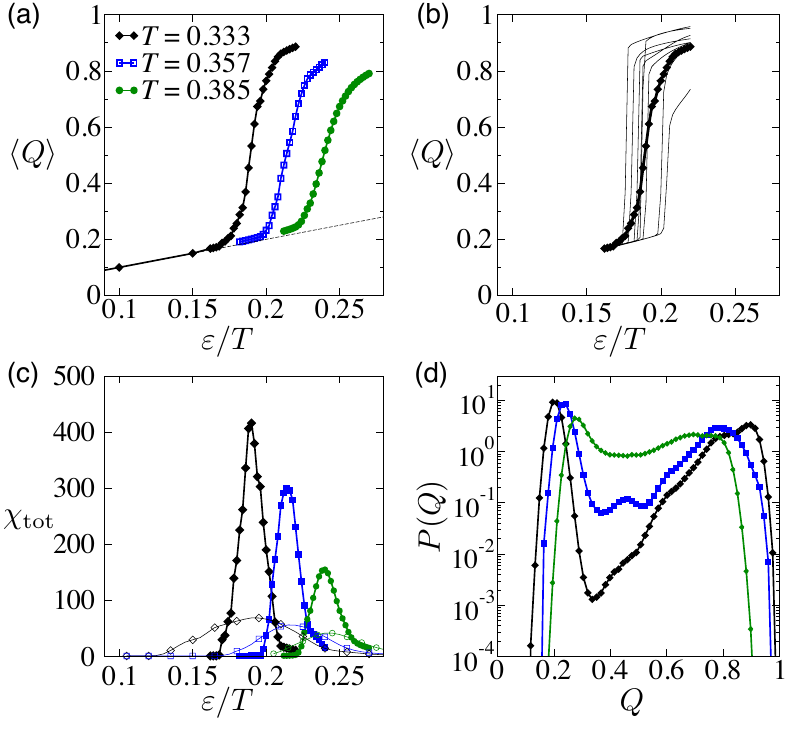}
\caption{Results for $T=T'$.
(a)~Mean overlap $\langle Q \rangle$ as a function of $\eps$, showing an increasingly sharp jump as $T$ is reduced.  
The dashed line is $\langle Q \rangle = \eps/T$, which gives
the linear response behaviour around $\eps=0$.  The system size is $N=16^3$.
Colors and symbols have the same meaning in all panels.
(b)~Average overlap for $T=0.333$, as well as thermal averages $\langle Q \rangle_{\CC'}$ for eight representative
reference configurations.
(c)~Total susceptibility $\chi_{\rm tot}=N\langle \delta Q^2\rangle$.  Solid symbols
are for a system of size $N=16^3$ while open symbols are for $N=8^3$, at the same temperatures. 
(d)~Distributions $P(Q)$ for $N=16^3$, at the values of $\eps$ which maximise $\chi_{\rm tot}$, showing increased bimodality on reducing temperature.
}
\label{fig:bbp}
\end{figure}

\begin{figure}
\includegraphics[width=8.5cm]{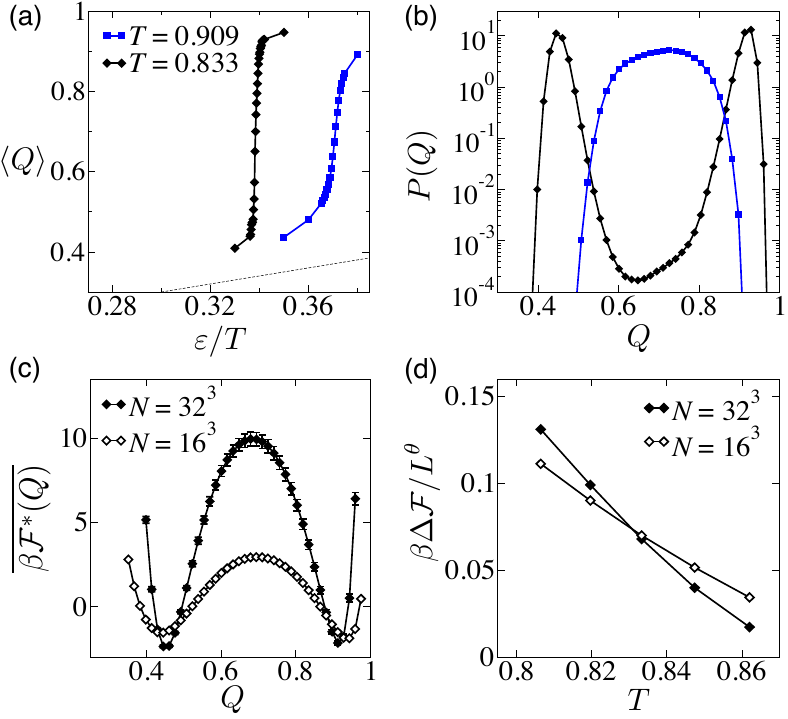}
\caption{Results for $T'=0.25$. 
(a)~The mean overlap shows a sharp crossover as $\eps$ is increased.  The system size is $N=32^3$. The dashed line is the linear 
response relation $\langle Q \rangle = \eps/T$.
(b)~Overlap distribution $P(Q)$, evaluated at $\eps=\eps^*$; the symbols and colors indicate the same temperatures shown in~(a).
(c)~Free energy $\overline{\beta{\cal F}^*(Q)}$ for $T=0.833$, showing that the interfacial free
energy cost (barrier between the two minima) increases strongly with system size.
Error bars show standard errors for $N=32^{3}$; numerical uncertainties for $N=16^{3}$ are comparable with system sizes.
(d)~Temperature dependence of the scaled interfacial free energy cost,  $\beta \Delta F/L^\theta$, from which we estimate
$T_c\approx0.83$.  
}
\label{fig:bp4}
\end{figure}

%
For $T'=0.25$, Fig.~\ref{fig:bp4}(a) shows a sharp jump in $\langle Q \rangle$ as $\eps$ increases, this time at a relatively high temperature, $T\approx 0.8$, for which simulations are more tractable.
This allows investigations of larger systems, up to $N=32^{3}$.  Fig.~\ref{fig:bp4}(b) shows the distribution of the overlap $P(Q)$, whose form changes from
 unimodal to bimodal  as the temperature is lowered.   
 
 To identify the critical point, we use finite-size scaling.  
However, the RFIM has unusual finite-size scaling properties: the transition is \emph{almost first order} and the distribution $P(Q)$ has two
well-separated peaks even at the critical point~\cite{vink2010}.  The order parameter exponent $\beta$ is also very close to zero~\cite{fytas2013}.  Together with the significant
numerical uncertainties in $P(Q)$, these two features lead to difficulties with
classical finite-size scaling based on universal cumulant ratios or order parameter distributions.  To address this problem, we 
follow~\cite{vink2010}: for each reference configuration
$\CC'$, we calculate the coupling $\eps^*_{\CC'}$ that maximises the variance $\langle \delta Q^2\rangle_{\CC'}$.
At this coupling, the distribution of $Q$ is typically bimodal, even above $T_c$~\cite{vink2010}.  We define a free energy for
this reference configuration ${\cal F}^*(Q)=-T\ln P_{\CC'}(Q)$.
This free energy has two minima, and the height of the barrier between them is the interfacial free energy cost between high- and low-overlap
states.  To obtain an average free energy cost, we calculate the average free energy $\overline{{\cal F}^*(Q)}$ from which we extract
the barrier height $\Delta \cal F$.

Fig.~\ref{fig:bp4}(c) shows the resulting free energy (scaled by $\beta=1/T$ to allow interpretation as a log-probability). 
In contrast to $P(Q)$, the numerical uncertainties in ${\cal F}^*(Q)$ are straightforward to estimate, and
relatively small.
For RFIM critical points, one expects a sigificant interfacial free energy in small systems
even for $T>T_c$, but $\lim_{L\to\infty}\Delta {\cal F}\to 0$ in that case.
For $T<T_c$, one expects a divergence
of the interfacial cost as
$\Delta {\cal F}\sim L^{d-1}$, as usual for first order phase transitions (here, $d=3$ is the spatial dimension).  
At the RFIM critical point, the interfacial free energy cost $\Delta {\cal F}$ scales as $L^\theta$~\cite{Fisher1986,vink2010}.  
In Fig.~\ref{fig:bp4}(d), 
we plot $\beta\Delta {\cal F}/L^\theta$ for two system sizes (we take $\theta=1.5$).  The interfacial costs are large and they grow with system size, Fig.~\ref{fig:bp4}(b);
they grow increasingly rapidly at low temperatures, Fig.~\ref{fig:bp4}(c). This presents strong evidence for the existence of a critical
point, whose properties are consistent with RFIM universality.  From the crossing point in Fig.~\ref{fig:bp4}(d),
we estimate 
the critical temperature
as $T_c\approx0.83$.

\begin{figure}
\includegraphics[width=8.5cm]{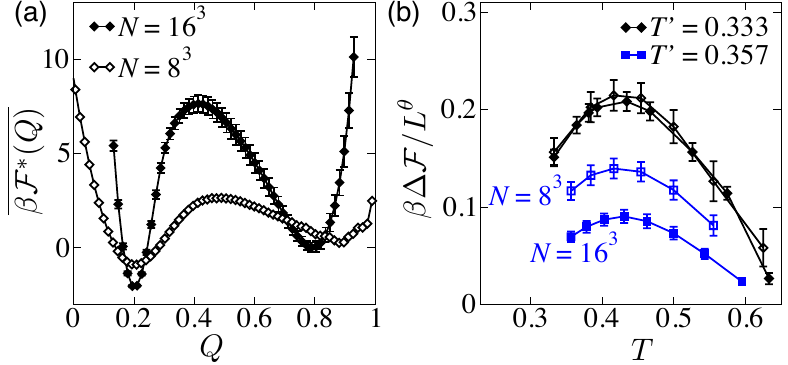}
\caption{(a)~Free energy $\overline{\beta {\cal F}^*(Q)}$  for $T=T'=0.333$ and system sizes $N=(8^3, 16^3)$.  The interfacial
cost increases strongly with system size.  The system is at coexistence: the difference in height between the two minima reflects their different widths.
(b)~Scaled interfacial costs $\beta\Delta {\cal F}/L^\theta$; filled symbols are $N=16^3$
and open symbols $N=8^3$.  The costs are non-monotonic in $T$; their finite-size scaling is discussed in the main text.
}
\label{fig:bbp-F}
\end{figure}

\emph{Interfacial costs for $T'=T$}.
Armed with these results for $T'=0.25$, we now return to the important case $T'=T$.  
Fig.~\ref{fig:bbp-F}(a) shows the free energy $\overline{{\cal F}^*(Q)}$,
at $T=T'=0.333$.  The interfacial cost
increases strongly with system size.   Fig.~\ref{fig:bbp-F}(b) shows the temperature-dependence of this cost.
For $T'=0.357$ and $T\geq T'$, the scaled free energy cost always decreases as the system size is increased from $N=8^{3}$ to
$N=16^{3}$.  This indicates that these systems are all above $T_c$.  We also note that these free energy costs are
non-monotonic in temperature  -- our interpretation of this unusual feature is that the high-overlap and low-overlap states have increasingly
similar structures as $T$ approaches $T'$, which tends to reduce the surface tension between them.  Moreover, for $T'=0.333$, the scaled
free energy costs for the two system sizes are very close to each other over a wide range of temperature $T$.  [This is in contrast to the clear
crossing of the curves in Fig.~\ref{fig:bp4}(d).]  Our interpretation of this last result is that all these systems are close to criticality.

\begin{figure}
\includegraphics[width=8.5cm]{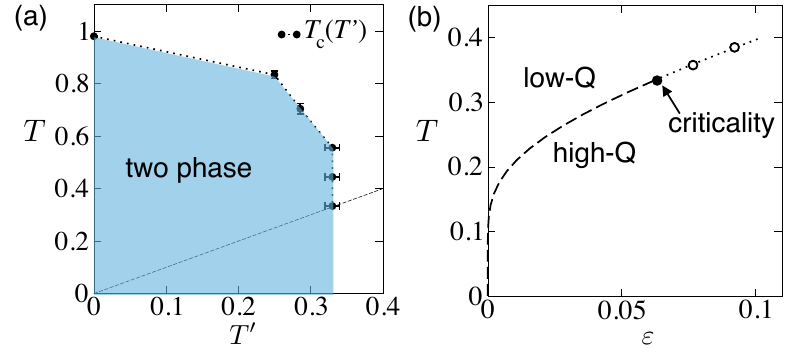}
\caption{(a) Estimated critical temperatures $T_c$ as a function of $T'$.  These critical points delimit a  two-phase region (shaded)
within which the Franz-Parisi potential $V(Q)$ includes a Maxwell construction, associated with phase coexistence.  The
line of critical points intersects the dashed line $T=T'$ at $T\approx 0.33$.
(b) Sketched phase diagram for the SPyM with $T'=T$.  The points indicate the susceptibility maxima from Fig.~\ref{fig:bbp}, with
the estimated critical point at $T\approx 0.33$.
Lines follow $\eps=A\ee^{-1/T}$ (see main text) with fit parameter $A=1.25$.  They indicate 
a first-order phase transition (dashed) and a susceptibility maximum (dotted).
}
\label{fig:TcT}
\end{figure}

\emph{Summary of phase behaviour} --
To illustrate the resulting scenario we consider the critical temperature $T_c$ as a function of the reference temperature $T'$.
For $T'=0$ the reference configuration $\CC'$ 
is in its ground state which has all spins up, and the system reduces to a SPyM in a magnetic field, for which there is known to be
an Ising critical point at $T_c=0.98$ \cite{Heringa1989,Turner2015}.  On increasing $T'$, we expect a line of critical points in the parameter space $(\eps,T,T')$,
all of which should be of RFIM type, except for the special (Ising) case $T'=0$.   The estimated critical ponts that we have found in this study are shown
in Fig.~\ref{fig:TcT}(a).  They separate a region of parameter space in which phase coexistence is possible (``two phase'') from a one-phase region
where the response to the coupling $\eps$ is smooth through the entire range of the overlap ($0<Q<1$).  The Franz-Parisi potential $V(Q)$
is strictly convex in the one-phase region but includes a linear segment (Maxwell construction) in the two-phase region.
(Evidence for the critical point with $T'=0.286$ is shown in Appendix B.) 

The dashed line in Fig.~\ref{fig:TcT}(a) indicates $T=T'$. It intersects the line of critical points at $T'\approx 0.33$. 
 When coexistence takes place at $T=T'$, the nature of the interfacial free energy cost is a subtle one, which has important implications for theories of the glass transition.
Since $\CC'$ is an equilibrium configuration of the isolated system, one expects the high-$Q$ state to have configurations $\CC$
that are also close to equilibrium.  Similarly, the low-$Q$ states are only weakly perturbed from equilibrium, so one expects $\CC$
to have close-to-equilibrium structures in that case too.  Thus, the phase coexistence that is illustrated in Fig.~\ref{fig:bbp-F} is taking
place between different states which have (statistically) very similar structures.  The interfacial cost associated with such states
plays a central role in random first order transition (RFOT) theory~\cite{Bouchaud2004}, because $\CC'$ may be viewed as a form of self-generated disorder.
However, the nature of these interfaces in finite-dimensional systems is not fully understood~\cite{Moore2006,Cammarota2009}.
As noted above, the reduction in interfacial cost as $T$ approaches $T'$ in Fig.~\ref{fig:bbp-F}(b) is
consistent with the idea that the high- and low-overlap states are becoming increasingly similar.  It also indicates
that the cost is primarily entropic and not energetic in origin (otherwise the probability cost would likely increase on cooling, as $\beta\Delta{\cal F} = \Delta E/T$).  

Still concentrating on the case $T=T'$,
the phase coexistence for $T<T_c$ reveals that the system
supports many metastable states, with a finite interfacial tension between them.
However, we emphasize that for $\eps=0$ the free energy of the SPyM is a simple analytic function of temperature and there
are no finite-temperature phase transitions.  
In general one expects a first-order transition at $\eps^*/T \approx s_c\Delta Q$ where $s_c$ is the configurational entropy density and $\Delta Q$ the jump
in the overlap at $\eps^*$~\cite{Berthier2014entropy}. For plaquette models one expects $s_c$ to be of the same order as the total entropy $s=\ee^{-1/T}( 1/T -1)$,
and $\Delta Q\sim1$, which gives an estimate of the critical coupling, $\eps^* \sim \ee^{-1/T}$.
In Fig.~\ref{fig:TcT}(b), we sketch the expected
phase behavior of the SPyM, including
three data points that indicate the locus of $\eps^*$ as $T$ is varied, holding $T'=T$ (recall Fig.~\ref{fig:bbp}(a)).  
Given the limited data, the fitted line  $\eps^*\sim \ee^{-1/T}$ should be regarded only as a qualitative prediction for the low temperature
 behaviour, but we emphasise that the first-order line can meet the $\eps=0$ axis only at $T=0$.
{(
Similar scaling is 
found for annealed coupling between plaquette models,
although the phase transitions are of Ising type (not RFIM) in that case~\cite{Garrahan2014,Turner2015}.)}

\emph{Discussion} --
The behaviour of the SPyM is characteristic of the DF perspective of the glass transition (absence of a thermodynamic transition, effective kinetic constraints, facilitated relaxation and heterogeneous dynamics).  We have shown, nevertheless, that when coupled to a quenched reference state, the SPyM displays thermodynamic overlap transitions in the RFIM class, 
as expected from mean-field calculations and RFOT theory.
It follows that quenched overlap transitions in atomistic simulations \cite{Berthier2015} are not incompatible with DF theory; also transitions occurring at $T,\eps>0$ do not imply any finite-temperature singularities at $\eps=0$ (recall Fig.~\ref{fig:TcT}(b)).  
More generally, and taken together with other recent studies of plaquette models~\cite{Foini2012,Cammarota2012patch,Franz2015}, 
the presence of overlap transitions in the SPyM shows how these models can be consistent at the same time with predictions of both DF and RFOT theory,
offering a link between these scenarios.

To forge such a link, the crucial question is whether (and how) phase transitions for $\eps,T>0$ are related to
the unbiased ($\eps=0$) properties of the system.  Given the first-order transition line in the $\eps$-$T$ plane, we expect
a nucleation-like mechanism for dynamical relaxation near this line, with an initially high-overlap state decaying to a low-overlap one 
via nucleation and growth of a droplet of the low-overlap phase.  This situation is close to the RFOT picture described in~\cite{Bouchaud2004}: for $\eps<\eps^*$ a nucleation argument yields a free energy barrier
$\beta\Delta F \sim (\eps^*-\eps) R_c^d \sim (\eps^*-\eps)^{1-d}$ where $R_c\sim1/(\eps^*-\eps)$ is the size of the critical nucleus, which diverges at the first order transition line.
However, for $\eps=0$ the natural picture is that associated with DF: 
point-like defects facilitate co-operative rearrangements over a length scale $\xi \sim \ee^{1/(Td_{\rm f})}$ where $d_{\rm f}$ is a scaling
exponent~\cite{Garrahan2002}. The free energy barrier for these processes scales as $\beta \Delta F \sim (1/T) \ln \xi \sim (1/T)^2$.  
Bridging between the two regimes
 $\eps\sim \eps^*$ and $\eps\simeq 0$ remains a challenge.  
 However, our results suggest that such an interpolation might be
 a way to connect the defect-mediated dynamics of DF theory to the nucleation picture predicted by RFOT and mean-field theories.

RLJ thanks Ludovic Berthier, Gilles Tarjus, and Mike Moore for helpful discussions.  RLJ was supported by EPSRC grant No.\ EP/I003797/1 and JPG by 
EPSRC Grant No.\ EP/K01773X/1
\bibliography{spym}%

\begin{appendix}

\newcommand{\refAnnTPM}{16}

\section{Numerical methods}

\subsection{Thermal averages}

Our methods for simulating the SPyM follow closely those of [\refAnnTPM].  We start by using the continuous time Monte Carlo (CTMC)
method of~\cite{Bortz1975} to simulate the model, using Glauber rates for spin flips that take into account only the
uncoupled energy $E_0$.  We modify this algorithm slightly, in order to sample ensembles where the distribution of the configuration $\CC$ is
\begin{equation}
p_{\rm b}(\CC|\CC') \propto \ee^{-E_0(\CC)/T + \eps N Q(\CC,\CC')/T } \frac{1}{ b(Q(\CC,\CC')) } .
\label{equ:pb}
\end{equation}
Here $b(Q)$ is a bias function that is chosen to promote rapid sampling of a wide range of $Q$-values in a single simulation~~\cite{bruce2003}.  The 
construction of a suitable choice for $b$ is discussed below.  

To sample the distribution (\ref{equ:pb}) within the CTMC dynamics, we incorporate an additional
accept/reject step for each move, using a Metropolis criterion 
\begin{equation}
P_{\rm acc}=\mathrm{min}\left\{1,\exp\left[{(\eps N/T) (Q_{\rm new} - Q_{\rm old})}\right]
\frac{b(Q_{\rm old})}{b(Q_{\rm new})}\right\} .
\end{equation}
It is easily verified that this Monte Carlo algorithm converges to a steady state whose a distribution is given by (\ref{equ:pb}).
The desired Boltzmann distribution is then obtained by a reweighting $p(\CC|\CC') \propto p_{\rm b}(\CC|\CC') \cdot b(Q(\CC,\CC'))$.

We use an automated algorithm for choosing and updating the bias $b(Q)$.
The method operates for a fixed reference configuration $\CC'$, so the final (quenched) average over $\CC'$ simply
involves running the algorithm many times for a range of different reference configurations.

We focus on the biased distribution of the overlap $P_{\rm b}(Q|\CC')$ that is obtained from $p_{\rm b}(\CC|\CC')$.  Clearly, 
the biased distribution is $P_{\rm b}(Q|\CC')  \propto P_{\CC'}(Q) /b(Q) $ where  $P_{\CC'}(Q)$ is the unbiased overlap distribution
 discussed in the main text.
For systems close to first-order phase transitions, there is a long timescale
associated with spontaneous transitions between the phases. The time required for such a transition to occur spontaneously 
scales with the depth of the minimum in $P_{\CC'}(Q)$
and makes it difficult to assess the precise location of the phase transition (due to hysteresis effects).
The purpose of the bias function $b(Q)$ is to smooth out any deep minima in $P(Q)$, so that the distribution $P_{\rm b}(Q)$ is almost flat over the
range of $Q$ that is of interest~\cite{bruce2003}.  That is we choose, $b(Q) \approx P_{\CC'}(Q)$.

In practice, we start at a relatively high temperature $T$ for which sampling is easy and we estimate the value of $\eps$ 
such that $P_{\CC'}(Q)$ has maximal variance.  We use the MC dynamics with $b=1$ to collect $N_s$ representative
configurations $\CC^r$  with $r=1,2,\dots,N_s$.  We take $N_s$ in the range $10^3-10^4$:  samples are not all
independent from each other but the sampling runs are long enough that the system fully decorrelates within each run. 
For each sample, we store the overlap $Q_r=Q(\CC^r,\CC')$ and the energy $E_{r}=E_0(\CC^r)$.  
This provides an estimate for a suitable bias potential for further simulations at this state point:
\begin{equation}
b(Q) \propto \sum_r \delta_{Q,Q_r}
\label{equ:bm-emp}
\end{equation}
where $\delta_{Q,Q'}$ is the Kronecker delta.  Hence $b(Q)$ is the \emph{empirical distribution} of $Q$.  In practice, Eq~(\ref{equ:bm-emp})
can be used only in the range of $Q$ for which reliable data are available, and we take $b=\mathrm{const.}$ outside this range.

Now reduce the temperature to $T-\Delta T$.  We also reduce the coupling $\eps$ to a value $\eps'$ whose choice will be described below.  
We can estimate $P_{\CC'}(Q)$ at this new state point as
\begin{equation}
P_{\CC'}(Q) \approx b_\Delta(Q) \propto \sum_r \delta_{Q,Q_r} \exp( -\Delta_J E_r + \Delta_\eps N Q_r)
\label{equ:bm-new}
\end{equation}
where $\Delta_J = 1/(T-\Delta T) - 1/T$ is the change in inverse temperature and $\Delta_\eps = \eps'/(T-\Delta T)-\eps/T$. 
The parameter $\eps'$ (or equivalently $\Delta_\eps$) is chosen to maximise
the variance of $b_\Delta(Q)$, so that the new state point remains close to the susceptibility maximum as the temperature is reduced.
The parameter $\Delta T$ must be chosen small enough that $b_{\Delta}(Q)$ provides a reasonably accurate estimate of $P_{\CC'}(Q)$ at
 the new state point: in practice this means that the exponential weights in (\ref{equ:bm-emp}) should not lead to to a sum dominated by just
 a few terms.
 
 Repeating this procedure allows the temperature to be reduced while $b(Q)$ is maintained close to $P_{\CC'}(Q)$.  This is the method used
 to obtain the strongly bimodal distributions $P(Q)$ in Figs 1d, 2b, etc. of the main text. 
 
  We note in passing that the Franz-Parisi potential
 may be obtained as 
 \begin{equation}
 V(Q)=T\lim_{N\to\infty} N^{-1} \overline{\ln P_{\CC'}(Q)}.
\end{equation}
This function is therefore directly accessible by our method.  It is related to the free energy ${\cal F}^*(Q)$ of the main text as
 \begin{equation}
 V(Q)=\lim_{N\to\infty} \left[ \frac{\overline{\mathcal{F}^*(Q)}}{N} + \eps^* Q \right].
\end{equation}

\subsection{Averages over reference configurations}

Plaquette models such as the SPyM have the useful feature that equilibrated reference configurations can be generated in a simple way by choosing
random defect positions at a density $1/(1+\ee^{1/T})$ and then constructing the relevant spin configuration as described in~[\refAnnTPM].  The procedure described in the
previous section is
then repeated for each reference configuration, leading to the averaged quantities shown in the main text.  We typically calculated averages over
 64 reference configurations, except in cases where fewer samples were sufficient to obtain convergence.

\subsection{Numerical parameters}

In the main text we quote all temperature to three significant figures.  The exact temperatures used were $T=1/2.6, 1/2.8, 1/3.0$
(Fig.~1); $T=1/1.1,1/1.2$ and $T'=1/4$ (Fig.~2).  In Fig.~4 there is a single data point with $T'=1/3.5 \approx 0.286$.

\section{Results for $T=0.286$}

\begin{figure}[h]
\includegraphics[width=5cm]{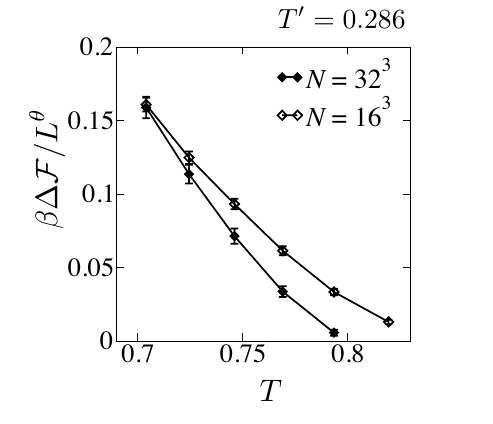}
\caption{Results for the scaled interfacial cost as a function of temperature $T$, with reference temperature $T'=0.286$.  Compare
Fig.~2d of the main text.}
\label{fig:bp35}
\end{figure}

Fig.~\ref{fig:bp35} shows the scaled interfacial cost in a system with reference temperature $T'=0.286$, comparing system sizes
$N=16^3,32^3$.  Compare Fig.~2d of the main text.  We infer that the critical temperature in this case is $T_c\approx 0.71$: the position of this 
critical point is shown in Fig.~4a of the main text. 

\end{appendix}

\end{document}